\documentclass[useAMS,usenatbib,referee]{article}

\def\bSig\mathbf{\Sigma}

\bibstyle{apalike}

\usepackage{graphicx}
\title{Estimation for network snowball sampling: Preventing pandemics}

\author{Steve Thompson  \\
Department of Statistics and Actuarial Science \\
Simon Fraser University, Burnaby, BC, Canada\\
thompson@sfu.ca}

\newlength{\normalparindent}
\raggedright
\begin{document}

%  This will produce the submission and review information that appears
%  right after the reference section.  Of course, it will be unknown when
%  you submit your paper, so you can either leave this out or put in 
%  sample dates (these will have no effect on the fate of your paper in the
%  review process!)

%\date{{\it Received October} 2007. {\it Revised February} 2008.  {\it
%Accepted March} 2008.}

\maketitle
\date{}

%  These options will count the number of pages and provide volume
%  and date information in the upper left hand corner of the top of the 
%  first page as in published papers.  The \pagerange command will only
%  work if you place the command \label{firstpage} near the beginning
%  of the document and \label{lastpage} at the end of the document, as we
%  have done in this template.

%  Again, putting a volume number and date is for your own amusement and
%  has no bearing on what actually happens to your paper!  

% \pagerange{\pageref{firstpage}--\pageref{lastpage}} 
% \volume{64}
% \pubyear{2008}
% \artmonth{December}

%  The \doi command is where the DOI for your paper would be placed should it
%  be published.  Again, if you make one up and stick it here, it means 
%  nothing!

% \doi{10.1111/j.1541-0420.2005.00454.x}

%  This label and the label ``lastpage'' are used by the \pagerange
%  command above to give the page range for the article.  You may have 
%  to process the document twice to get this to match up with what you 
%  expect.  When using the referee option, this will not count the pages
%  with tables and figures.  

\label{firstpage}

%  put the summary for your paper here

%  Please place your key words in alphabetical order, separated
%  by semicolons, with the first letter of the first word capitalized,
%  and a period at the end of the list.
%

\begin{abstract}
Snowball designs are the most natural of the network sampling
designs.  They have many desirable properties for sampling hidden and
hard-to reach populations.  They have been under-used in recent years
because simple design-based estimators and confidence intervals have
not been available for them.  The needed estimation methods are
supplied in this paper.  Snowball sampling methods and accurate
estimators with them are needed for sampling of the people exposed to
the animals from which new coronavirus outbreaks originate, and to
sample the animal populations to which they are exposed.  Accurate
estimates are needed to evaluate the effectiveness of interventions to
reduce the risk to the people exposed to the animals.  In this way the
frequencies of major outbreaks and pandemics can be reduced.  Snowball
designs are needed in studies of sexual and opioid networks through
which HIV can spread explosively, so that prevention intervention
methods can be developed, accurately assessed, and effectively
distributed.

\end{abstract}


%  As usual, the \maketitle command creates the title and author/affiliations
%  display 

\maketitle

%  If you are using the referee option, a new page, numbered page 1, will
%  start after the summary and keywords.  The page numbers thus count the
%  number of pages of your manuscript in the preferred submission style.
%  Remember, ``Normally, regular papers exceeding 25 pages and Reader Reaction 
%  papers exceeding 12 pages in (the preferred style) will be returned to 
%  the authors without review. The page limit includes acknowledgements, 
%  references, and appendices, but not tables and figures. The page count does 
%  not include the title page and abstract. A maximum of six (6) tables or 
%  figures combined is often required.''

%  You may now place the substance of your manuscript here.  Please use
%  the \section, \subsection, etc commands as described in the user guide.
%  Please use \label and \ref commands to cross-reference sections, equations,
%  tables, figures, etc.
%
%  Please DO NOT attempt to reformat the style of equation numbering!
%  For that matter, please do not attempt to redefine anything!

\section{Introduction}

\begin{figure}
 \centerline{\includegraphics[width = 1.0\textwidth]{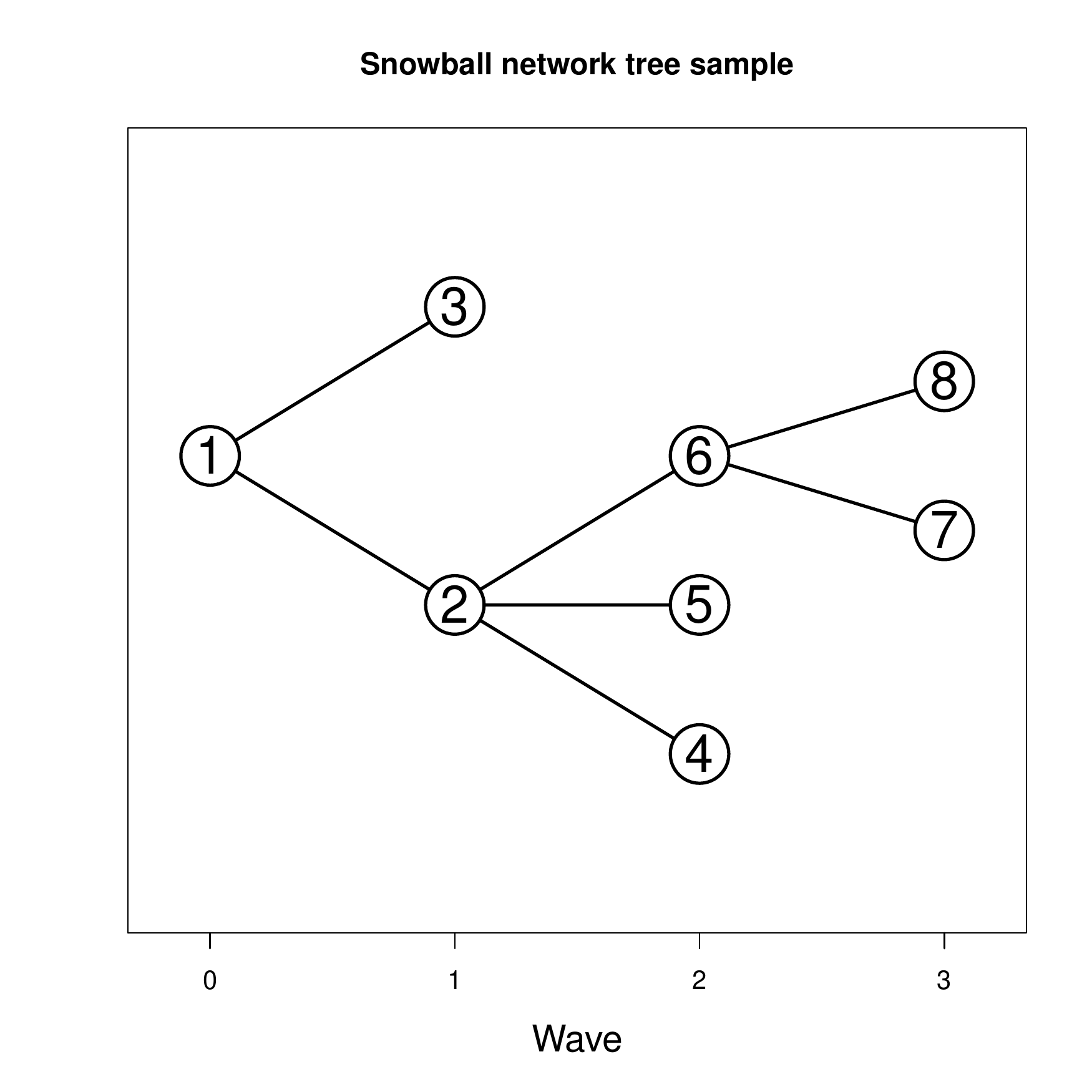}}
 \caption{A sample from a regular snowball design.  Repeat selections
   of the same person are not allowed.  This results in a network
   sample with tree structure.}
\label{fig:sbtree}
\end{figure}

\begin{figure}
 \centerline{\includegraphics[width = 1.0\textwidth]{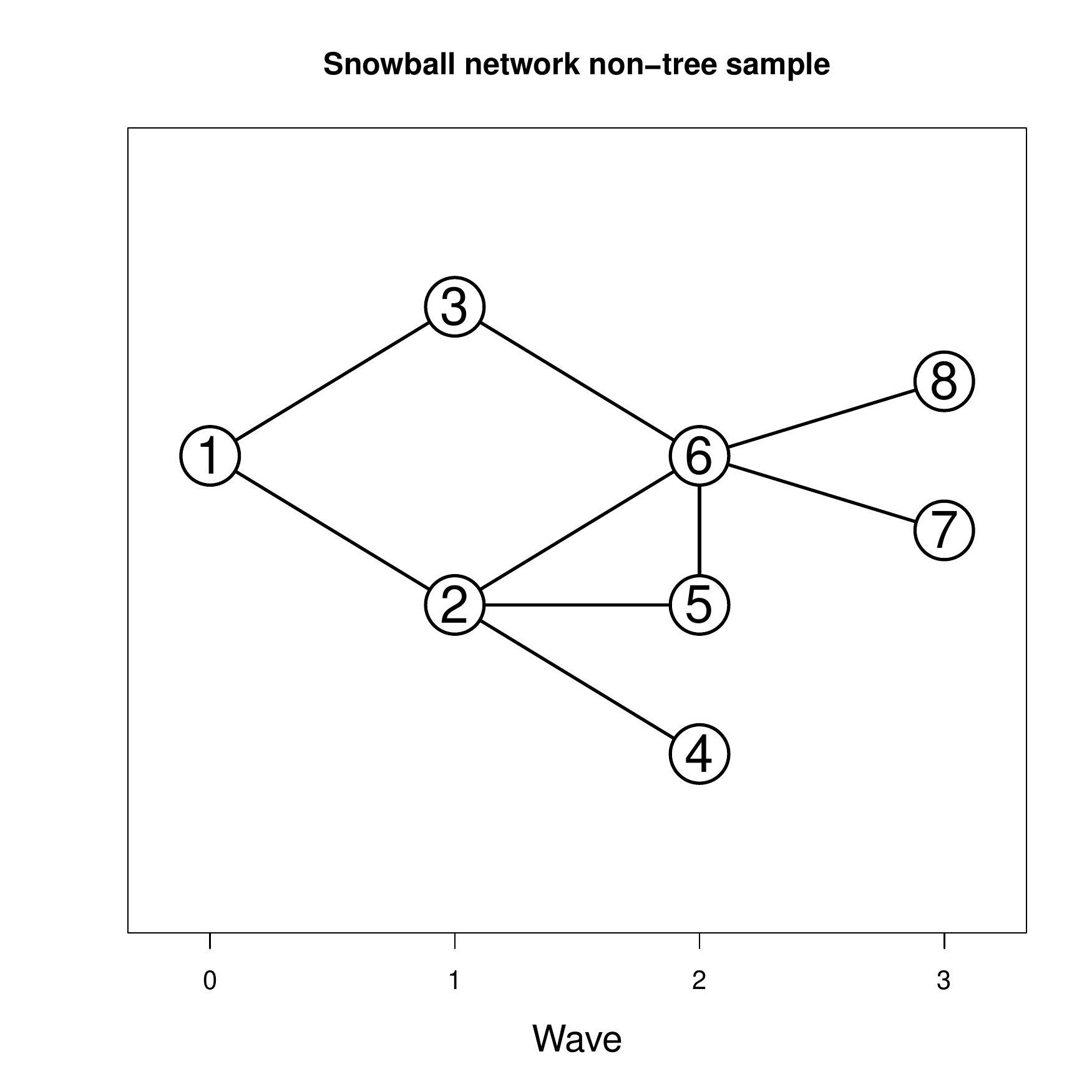}}
 \caption{A sample from a snowball design allowing repeat selections
   of the same person, but only by a different recruiter.  This
   results in a sample with network structure more general
   than a tree.}
\label{fig:sbnontree}
\end{figure}

In network sampling, social links are followed from one person to
another in selecting the sample of people from a subpopulation that
might otherwise be hard to reach.  The most natural of the network
designs is snowball sampling, in which any number of the links out
from each person may be followed.  So if a person in the sample has two
partners or contacts, either zero, one, or two of those contacts are
added to the sample, depending the selection probabilities of the links.
For another person in the sample with ten links to contacts, any
number from zero to ten links may be followed.  In turn, links out
from the new people in the sample are followed at the next wave.  In
this way the sample may grow fast or slowly, depending on the
link-tracing probabilities and also depending on the numbers and
configuration of the links in the social network of the target
population.

In some snowball surveys, a respondent is not allowed to recruit a
person who has already been recruited into the sample.  This results
in a network sample with tree structure (Figure 1).  In other cases it
may be natural to allow re-recruitments.  Usually the survey protocol
will not allow re-cruitment of a person by the same person that
recruited them earlier, but will allow re-recruitment of a sample
member by a different recruiter.  This results in samples with network
structure more general than a treee (Figure 2).  Note that most people
in the sample have roles as both recruitees and recruiters.

The nature of snowball sampling brings different people into the
sample with different probabilities.  To estimate the characteristics
of the target population, such as mean number of partners or
proportion of people with a characteristic such as a risk-related
behavior, the different inclusion probabilities for different sample
members should be taken into account.  A sample mean or sample
proportion gives each person equal weight and so results in a estimate
that is biased in comparison to the actual population mean or
proportion.  Unbiased estimators in this design-based sense have been
available for only simple forms of snowball sampling, such as a
complete one-wave snowball sample where the inclusion probabilities
for the initial sample are known.  The problem with more general
snowball samples, with many waves and link-tracing probabilities less than
one, is that the inclusion probabilities are not known.  In fact they
can not be exactly calculated from the sample data because they depend on
the population network outside the sample as well as within it, as
well as depending on the sampling design.

This inference problem has kept snowball sampling designs from being
more widely used than they should be.  In this paper we present new
estimators for key population characteristics from snowball samples.
The method is based on estimating the unit inclusion probabilities
from the sample data in interaction with the network sampling design.
The method is design-based and data-based.  It does not assume a
statistical form of model for the population network.

Snowball designs are currently being used to collect data on people
with exposure to animals, in order to understand and prevent the
emergence of novel species of the coronavirus family that jump from
animals to humans and lead to emergence of epidemics or pandemics
every few years.  Snowball designs are used  in some
studies of key populations at high risk for HIV, such as sex workers,
clients of sex workers, and networks of people who mis-use opioids and
other drugs that are in some cases injected.  Contact-tracing designs
of people exposed to sexually transmitted diseases or exposed to
someone with the new disease covid-19, where the purpose is primarily
to find cases and make interventions, rather than inference, also are
of snowball form.

Because the of the importance of the snowball studies of in relation
to emergence of outbreaks of new coronavirus species in humans, and
the fact that those studies are recent and not so well known, that
example will be described at some length next.  Subsequent sections
will describe the method proposed here and evaluate it in relation to
alternatives.

Snowball sampling designs have many important uses in the real world.
In this paper we will focus on three.  One concerns the prevention of
future pandemics caused by emergence of new species in the coronavirus
family.  The second, related to the first, involves spatial adaptive
designs for animal species with uneven distributions.  The third
involves studies of sexual and opioid networks with high risk
of HIV, the human immunodeficiency virus.

\subsection{Outbreaks of new species of coronavirus in humans}

Virus species of the coronavirus family are adept at transfering from one
mammal host species to another, adapting to the new host environment
through genetic mutations and recombinations
(\cite{latinne2020origin}).  These can include transfers from bats,
civets, raccoon dogs, pangolins and other species to humans
(\cite{graham2010recombination}, \cite{tang2020origin},
\cite{andersen2020proximal}, \cite{montgomery2020covid}).  Bat species
are considered likely to be the main reservoir for viruses in the the
coronavirus family.  These transfers to humans are moderately common.
Most of these die out as the person infected recovered or dies, or
they infect just a few people close to that person before dying out
(\cite{zhou2020pneumonia}).  But a small proportion of these transfer
viruses catch on and become major outbreaks.

Since the start of the 21st Century there have been three major
outbreaks of coronavirus species in humans.  These are SARS ( the
species SARS-CoV-1) in 2002, the Middle East Respiratory Syndrome
(MERS), in 2012, and covid-19 (caused by the coronavirus species
SARS-CoV-2), in 2019, and at the time of this writing a worldwide
pandemic.  More such outbreaks and pandemics are widely expected by
virologists and epidemiologists unless some interventions can make
conditions less favorable to them.

\cite{li2019human}  used snowball sampling to obtain a sample across three
provinces in the south of China of 1596 people at-risk through their
exposure to animals.  Eight study regions were selected in areas known
to have diverse coronaviruses in bat populations roosting close
to human dwellings.  The study targeted ``human populations that are
highly exposed to bats and other wildlife, including people who visit
or work around bat caves, work in local live animal markets, raise
animals, or are involved in wildlife trade (e.g., wild animal harvest,
trade, transportation, and preparation), as identified by previous
exploratory ethnographic interviews.''

Sample members were given a questionnaire about risk-related and
protective behaviors and their experiences with flu-like and SARS-like
symptoms, and were given antibody tests for evidence of past infection
with bat coronavirus.  The blood tests identified 9 individuals
(0.6\%) who were positive for bat coronaviruses.  73 (5\%) reported
symptoms fever with cough and shortness of breath or difficulty
breathing within the last 12 months, and 227 (14\%) reported fever
with muscle aches, cough, or sore throat symptoms within the last 12
months.

When asked about protective measures, among the 502 participants
who purchased animals from wet markets in the past 12 months 194,
(39\%) had taken measures such as washing hands, purchasing live
animals less often , or purchasing meat at supermarkets instead of
live animal markets. Only 7 (1\%) had considered wearing a mask
while visiting the markets, with the same number having considered
wearing gloves.

Each reported percentage is the unweighted sample mean
(divided by 100). Much more accurate estimates for each of those
quantities can be calculated using the method proposed in this paper.

The need for accurate estimates becomes acute when we consider
evaluating the effectiveness of an intervention.  Suppose an
intervention program is introduced to a community to increase the
wearing of gloves when working in the markets.  We need to compare
wearing gloves in the community before and after the intervention, or 
with the rate in a community without the intervention program.  If the
estimates are highly biased or erratic in their statistical
properties, then the comparisons may be in error and lead to
choosing the less effective intervention for roll-out on a larger
scale.

A number of studies have sampled bats from their natural habitaats,
tested them for coronavirus species, and through phylogenetic analyses
established their relationship to species that have infected or can
infect humans (\cite{ge2013isolation}, \cite{li2005bats},
\cite{wang2006review}, \cite{wang2019viruses}, \cite{olival2017host},
\cite{olival2017erratum}, \cite{hu2017discovery},
\cite{corman2015evidence}).

%\cite{huong2020coronavirus}

\cite{huong2020journal.pone.0237129}  sampled the animals, rather than the
humans, along a supply chain in the wildlife trade in Vietnam.  The
animals on their way to market and restaurants were tested for
infection with bat coronavirus species.  They found that the infection
rate increased along the supply chain.  For example, field rats for food, in
the possession of the trappers who catch them wild in the field, had a
21\% infection rate.  For rats in the markets, the infection
rate was 32\%.  At the restaurants that served wild animal dishes, the
infection rate in the rats was 56\%.  The presumed reason for the
amplification of infection rate along the supply chain was
overcrowding in the cages the animals were transported and held in, so
that infection rapidly spread between the animals.

If the infection rate of the animals is higher at a point in the
supply chain, then the risk to the people who work with them will be
higher there.

The way to prevent future outbreaks and pandemics of emerging species of
coronavirus in humans is to decrease the frequency of occurrences of
transfers of coronavirises from animals to people.  Most of those
transfer strains die out before producing any significant outbreak,
but some small proportion of them will catch on to produce serious
outbreaks as well as new pandemics.  If there are less frequent
transfers, there should be correspondingly less frequent outbreaks
and pandemics.

So far in the 21st Century the serious outbreaks have come 7 and 10
years apart, so the next one could be anticipated to arrive not far in
the future.  If through interventions the frequency of transfers
could be reduced to one-third its current value, then the next outbreak
instead of 7 or 10 years, might be 21 or 30 years in the future.  The
anticipated outbreaks before then would have been prevented.

Potential interventions suggested by the currently available studies
would include programs to increase the use of protective measures
like washing hands and use of gloves and masks in working with
animals.  To reduce the amplification of infection rates along the
supply chain, safer and more humane transfer cages would need to be
designed and required.

Sampling both of the people exposed to animals and of the animals they
are potentially exposed to will be required to accurately assess the
effectiveness of different intervention strategies and select thereby
the best interventions to make.  The key missing tool up to this point
has been an accurate, robust estimation method for snowball sampling
designs.  The purpose of this paper is to present such a method.

\subsection{Adaptive spatial designs for animal species}

In addition to snowball sampling of the people exposed to animals, we
need to sample the animals they are exposed to, in order to estimate
the prevalence of coronavirus infection in those animals, and to
obtain the genetic sequences of those virus.  In some cases the
sampling of the animals will be straightforward, using conventional
sampling design features such as random sampling without replacement
and stratification.  This may be the the case in markets and wildlife
farms.  If we need to sample wild animals living nearby the people,
such as bats with natural roosting sites in trees, adaptive spatial
designs could be useful.

The adaptive designs work by starting with an initial conventional
sample, such as a random sample of spatial units.  Whenever a
significant number of animals is found in a sample unit, its
neighboring units are added to the sample.  If any of those has
significance abundance, its neighbors in turn are added, and so on
(\cite{thompson1990adaptive}).  In (\cite{thompson2006aws}, which was
first about network designs for estimating prevalence of risk
behaviors for HIV, it is shown that any of the spatially adaptive
design can be re-framed as a network sampling problem (see also
\cite{thompson2011adaptive}.  In (\cite{thompson2006aws} this was done
for a spatially uneven population of wintering waterfowl, and proved
very effective there.  In the same way the network sampling estimators
proposed in the present paper can be used for adaptive spatial
sampling designs and have some advantages in terms over previous
estimators for adaptive sampling designs in terms of simplicity and
flexibility.

The re-framing of an adaptive spatial design as network sampling works
as follows.  In the spatial setting supposse for simplicity the
sampling units are square plots covering the study region.  Convert
each square plot to a circle representing a node in a network.  If
plot $i$ has animals in it, then if we observe $i$ it will lead us to
any of its neighbors.  So for a neighbor $j$ draw an arrow from $i$ to
$j$.  If unit $j$ has animals in it, then if it was observed first it
would lead us to $i$, so draw an arrow from $j$ to $i$.  With arrows
in both directions, the link between $i$ and $j$ is symmetric, so it
can also be drawn as a simple line.  Now suppose unit $j$  has no
animal in it.  Then it will not lead us to its neighbor $i$.  So there
is an arrow from $i$ to $j$ but not from $j$ to $i$.  So the link is
directional here, and we get a network that is partially directional.
The estimation method work fine with the directional links in the
network.  This re-framing is illustrated in the figures of
\cite{thompson2006aws} and \cite{thompson2011adaptive}.

\subsection{HIV-at-risk sexual networks, opioid networks}

With the HIV epidemic the key populations at risk include people who
inject drugs non-medically, people who sell or buy sex, or trade it for
drugs or tangible goods, and men who have sex with men.  To study
these people and try to alleviate the epidemic, ethnographers and
health outreach workers used link-tracing methods.  In the 1980s and
1990s these were usually in the form of snowball designs.

Snowball sampling methods were used in \cite{potterat1999network} to
obtain data on the entire network and people of a population at risk
for heterosexual and drug-related spread of HIV.  In the study,
investigators endeavoured to find the individuals at both ends of every
relevant relationship.  Because of the relative geographic isolation
of the high-risk population, they were able to obtain essentially the
entire population and its network structure.  This population data set
has been invaluable to studies of network sampling and estimation
methodologies.  It is used in this paper for the simulations to
evaluate the effectiveness of the proposed estimators for snowball
network designs in comparison with other estimation methods.

Snowball sampling designs, the most freely-branching of the network
designs, are useful for studying the sexual and injection-related
networks in which HIV spreads.   \cite{peters2016hiv} and
\cite{campbell2017detailed} report on a study in which contact
tracing was used after an HIV outbreak associated with opioid misuse
in Indiana had spread rapidly. The traced network together with
phylogenetic data from sequencing of virus strains was used to
determine where the outbreak started and how it spread.  

Snowball designs are generally the preferred network sampling method
for bringing interventions to benefit a population.  Contact tracing
for sexually transmitted diseases has long used snowball sampling.  An
individual who tests positive for the disease is asked to identify all
their recent sexual partners.  Investigators attempt to find each of
those partners, inform them of their potential exposure, test then for
the infection and, if the test is positive, to treat them to cure the
infection and then in turn try to trace all their partners and do the
same.  In the current coronavirus pandemic, contact tracing of
individuals who test positive involves finding all their recent
contacts and at minimum to advise those individuals to self-isolate
for a period.  And in the more thorough versions, the contacts when
found are also tested and, if a test is positive, that person's
contacts are traced as well.

For HIV there is currently no cure or vaccine available.  There are
effective interventions such as increasing use of condoms and
targeting condom use to the early period in any new relationship.  In
\cite{thompson2017adaptive} it is shown that distributing such an
intervention program to some proportion of the population using a
snowball network design is highly effective in comparison to the same
intervention distributed to the same proportion of the population by
other methods.

\subsection{Snowball sampling}

Early uses of network sampling for hard-to-reach populations typically
used snowball sampling methods.  Ethnographers studying drug users and
other hidden populations wanted to know as many of the relationships
in the community so tried to follow every referral to social
partners.  The data was summarized by
unweighted sample means and proportions [\cite{spreen1992rare},
\cite{heckathorn1997respondent}, and \cite{thompson2002adaptive}].
Design-based estimates of population values from relatively simple
network sampling designs were obtained by \cite{birnbaum1965design}
and \cite{frank1977survey}, \cite{frank1994estimating}, and
\cite{birnbaum1965design}.

\cite{snijders1992estimation}  reviewed the literature on snowball
sampling for inference about population values. The estimates at that
time were limited to the more simple snowball designs, such as
one-wave designs with symmetric.   He concluded that
snowball sampling is more suited for inference about the links than
for inference about the nodes.
\cite{doi:10.1111/j.1467-9531.2011.01244.x} 
\cite{handcock2011comment} look at different uses of the
term ``snowball sampling''.

Model-based methods can work well for snowball designs but, in
addition to requiring a stochastic network model for the population,
may be limited too specific forms of snowball designs, such as ones in
which all links are traced out to a certain number of waves
(\cite{chow2003bayesian}, \cite{chow2003estimation}, \cite{thompson2000model}).

\cite{pattison2013conditional} assume an exponential random graph
model and describe a conditional estimation strategy for estimating
parameters of the model, with attention to computational efficiency so
that large samples can be handled.  The focus is on identifying
structural characteristics of the population network, such as stars
and triangles of various types..  Values of
nodes, such as whether an animal is infected or the person handling it
wears gloves, are not usually included in this type of model.
The method appears to work well for its intended purposes.  
  \cite{handcock2007modeling} provide an informative discussion of
  design characteristics in relation to model-based inferences for
  network samples.  
  \cite{handcock2010modeling}  describe a Bayesian model based
  approach for inference from snowball samples.  The method works well
  but is limited to specific types of snowball designs, such as those
  where every like is followed out to some wave.
  The main limitation of the model-based estimation methods for
  snowball sampling may be to scale-up, because the Markov Chain Monte
  Carlo method required in the computations get slow as sample size
  increases.  
\cite{atkinson2001accessing} review the methodology for snowball
sampling and conclude that snowball sampling is highly effective for
finding members of a hidden population but that statistical inference
from snowball samples was at the time of their writing problematical.

\cite{thompson2006aws} gives a design-based strategy for snowball
network designs with estimation based on one or another simple, if
inefficient, initial estimator.  That estimate is improved using the
Rap-Blackwell method, which finds the conditional expectation of the
initial estimator given the minimal sufficient statistics.  In the
network designs considered the minimal sufficient statistics is the
set of distinct units in the sample, not counting repeat selections
and not distinguishing order of selection.  The estimates obtained are
exactly unbiased and are very efficient, having low variance because
of the Rao-Blackwell improvement.  However, the method requires
control over the initial design that may not be achieved in practice
with hard-to-reach populations, and because of the Markov Chain Monte
Carlo needed to do the Rao-Blackwell computations, may be limited in
terms of scale up to large sample sizes.

In contrast to snowball sampling, 
 the methodology of
Respondent-Driven-Sampling (\cite{heckathorn1997respondent}), limited
the number of contacts a respondent could refer to a small number,
typically 3.  Estimators for these designs based on random walk theory
and assumptions of Markov transitions in the sampling between values
of attribute variables of respondents were given in
\cite{salganik2004sampling}, \cite{heckathorn20076}, and
\cite{volz2008probability}. In a random walk design only one link is
traced from the currently selected person, and the sampling is done
with replacement.  If a random walk is run
in a network consisting of a single connected component, the long term
frequency of inclusion of node $i$ is proportional to $d_i$.
Variations on these early approaches, still relying on the assumption
of a random walk design or Markov transitions include
\cite{gile2011improved}, \cite{baraff2016estimating}, and
\cite{rohe2019critical}.

Among network sampling designs, a snowball design is at the opposite
end of the spectrum from a random walk design.  A freely-branching
snowball design allows unlimited branching, up to however many links
are available, and is usually carried out without-replacement.  For
these reasons the estimators based on the random walk assumption have
never been recommended for snowball designs, nor should they be.

The idea of the new estimation method is very simple.  We run a sampling process
similar to the actual survey design on the sample network data and use
the inclusion frequencies in the sampling process to estimate the
inclusion probability of each sample unit in the real-world sampling.  With the
estimated inclusion probabilities, well-established sampling inference
methods can be used for population estimates and confidence
intervals.   Two approaches to the re-sampling are
repeated re-samples and a Markov chain resampling process.  For the
simulations we use the resampling process because it is
computationally much faster.

The estimation method and why it works are described more exactly in
the Methods section.  

\section{Methods}

\subsection{The new estimation method}

In traditional survey sampling with unequal probabilities of inclusion
for different people, typical estimators divide an observed value
$y_i$ for the $i$th person by the inclusion probability $\pi_i$ that
person.  A variable of interest $y_i$ can be binary, for example 1 if
the person tests positive for a virus and 0 otherwise, or can be more
generally quantitative, such as viral load.  The inverse-weighting
gives an unbiased or low-bias estimate of the population proportion or
mean of that variable.  In network surveys the inclusion probabilities
are unknown so they need to be estimated.

The estimators described in this report first estimate the inclusion
probability of each person in the sample by selecting many resamples
from the network sample data using a design that adheres in key
features to the actual survey design used to find the sample.  In
particular, the resampling design is a link-tracing design done
without-replacement and with branching, as was the original design.
The frequency $f_i$ with which an individual is included in the
resamples is used as an estimate of that person's inclusion
probability $\pi_i$.

What we do is select $T$ resamples $S_1, S_2, ..., S_T$ from the
sample network data.  There are two approaches to selecting the
sequence of resamples.  In the repeated-samples approach each resample
is selected independently from seeds and progresses step-by-step to
target resample size independently of every other resample, so we get
a collection of independent resamples.  in the sampling-process
approach each resample $S_t$ is selected from the resample $S_{t-1}$
just before it by randomly tracing a few links out and randomly
removing a few nodes from the previous resample and using a small rate
of re-seeding so we do not get locked out of any component by chance.
It is this Markov resampling process approach that we use for the
simulations in this paper because it is so computationally efficient.

For an individual $i$ in the original sample, there is a sequence of
zeros and ones $Z_{i1}, Z_{i2}, ..., Z_{iT}$, where $Z_t$ is 1 if that
person is included in resample $S_t$ and is 0 if the person is not
included in that resample.  The inclusion frequency for person $i$ is 
\begin{equation}
f_i =\frac{1}{T}\left( Z_{i1} + Z_{i2} + ... + Z_{iT} \right)
\label{eq:fi}
\end{equation}

  Individuals centrally
located in sample components tend to have high values of $f_i$.  That
is because there are more paths, and paths of higher probability,
leading the sample to those individuals.  Also, individuals in larger
components tend to have larger $f_i$ than individuals in smaller
components, so that the method is estimating inclusion probability of
an individual relative to all other sample units, not just those in
the same component or local area or the sample This is because of the
self-allocation of the branching design, even in the absence of
re-seeding, to areas of the social network having more links and
connected paths.

The estimator of the mean of a characteristic $y$ in the hidden
population is then 
\begin{equation}
\hat\mu_f = \frac{\sum (y_i/f_i)}{\sum (1/f_i)}
\label{eq:est}
\end{equation}
where each sum is over all the people in the sample.  If the actual
inclusion probabilities $\pi_i$ were known and replaced the $f_i$ in
Equation \ref{eq:est} we would have the generalized unequal
probability estimator $\hat\mu_\pi$ of \cite{brewer1963ratio}.
A simple variance estimator to go with $\hat\mu_f$ is
\begin{equation}
\widehat{\textrm var}(\hat\mu) =  \frac{1}{n(n-1)}  \sum_{s}
\left(\frac{ny_i/f_i}{\sum_s (1/f_i)} - \hat\mu_f\right)^2
\end{equation}
  An approximate
$1 - \alpha$ confidence interval is then calculated as $ \hat\mu \pm z
\sqrt{\widehat{\textrm var}(\hat\mu)}$, with $z$ the $1-\alpha/2$
quantile from the standard Normal distribution.

\vspace*{-8pt}

\begin{figure}
 \centerline{\includegraphics[width = 1.0\textwidth]{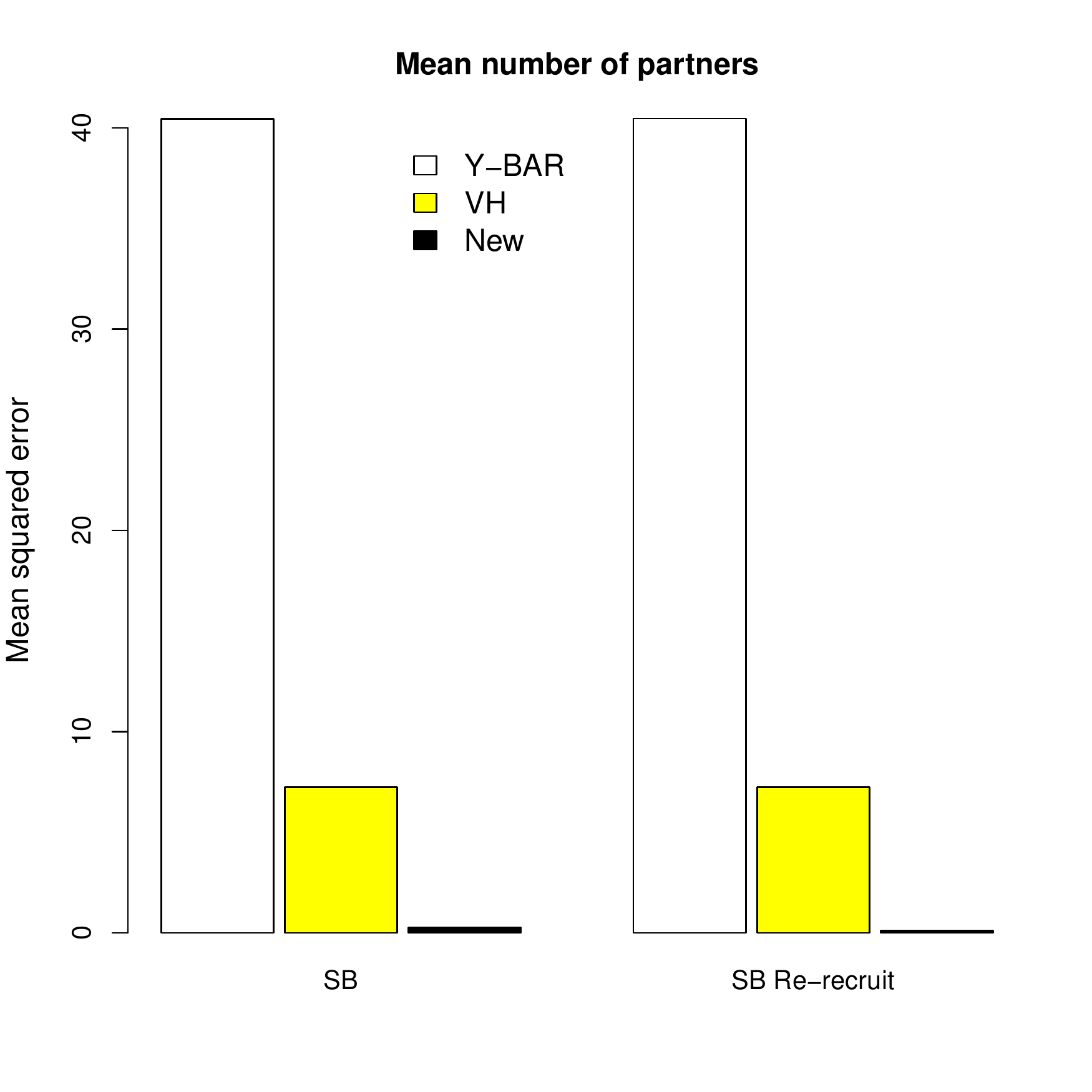}}
\caption{Mean squared error for three estimators of mean number of
  partners (mean degree), with a regular snowball design (left) and
  with a snowball design allowing re-recruitment of a person by
  different recruiters (right).  The three estimators are the sample
  mean (Y-BAR, white), weighting by reciprocal of self-reported degree
  (VH, yellow), and the new estimator (NEW, black).  Small MSE is
  good, so the new estimator is performing much the best.}
\label{fig:bardegree}
\end{figure}

\begin{figure}
 \centerline{\includegraphics[width =
   1.0\textwidth]{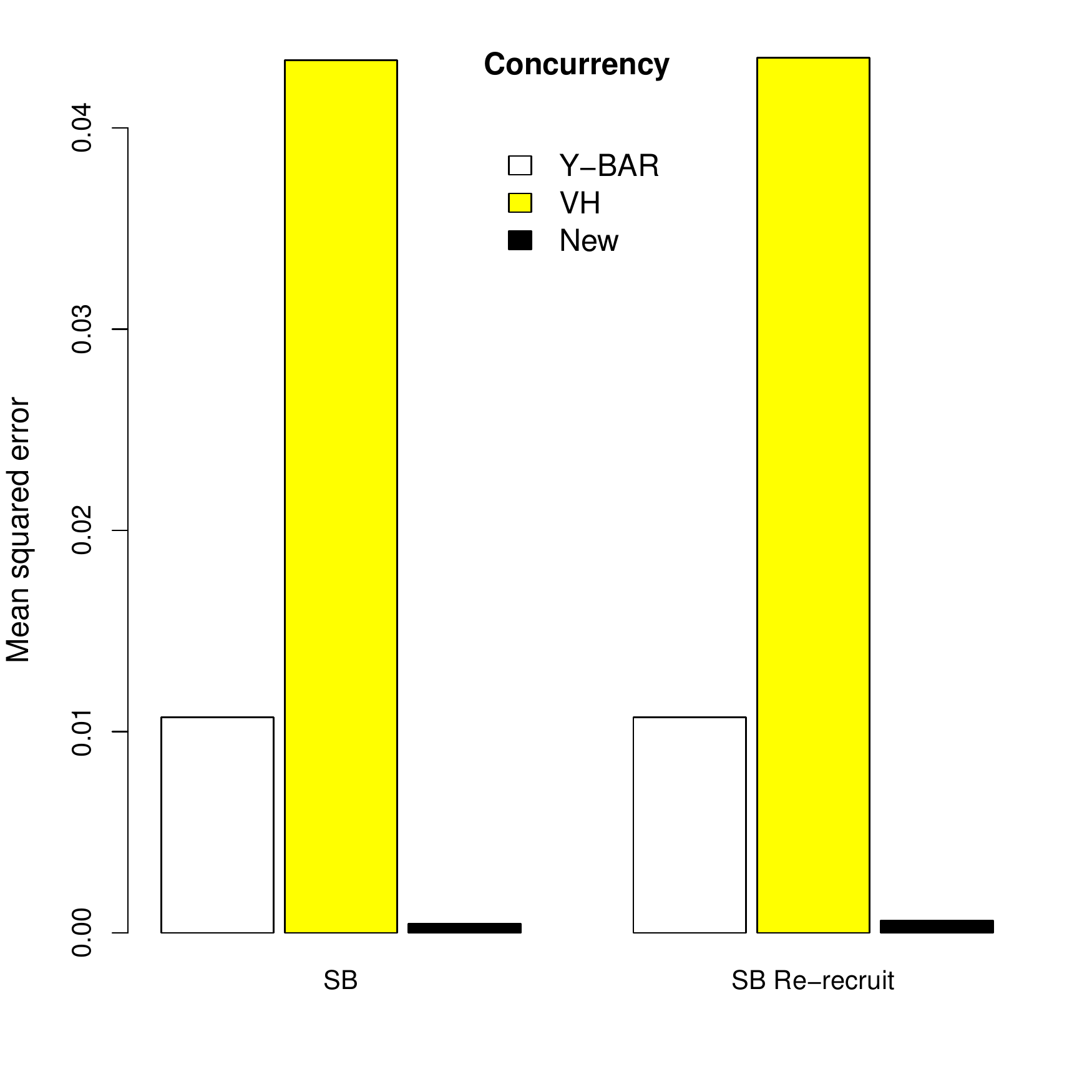}}
 \caption{Mean squared error for three estimators of concurrency, with a regular snowball design (left) and
  with a snowball design allowing re-recruitment of a person by
  different recruiters (right).  The three estimators are the sample
  mean (Y-BAR, white), weighting by reciprocal of self-reported degree
  (VH, yellow), and the new estimator (NEW, black).}
\label{fig:barconcurrency}
\end{figure}

\section{Results}

The new estimators are evaluated and compared with the current
estimates using the network data on the hidden population at risk for
HIV enumerated in the Colorado Springs study on the heterosexual
transmission of HIV [\cite{potterat1999network}], also known as the
Project 90 study.  The study was so thorough in finding every linked
person that it provides the most relevant network data set that can be
considered as an entire at-risk population for the purpose of
comparing sampling designs and estimators. In the simulations, the
population size is 5492, the number of people in the data set.  From
this population a sample of 1200 people is selected, using a snowball
design.  For resampling the target resample size is 400.  For each of
1000 samples of size 1200 each, 10,000 resamples were selected, each
resample of size about 400.  The population and the simulation methods
are described in more detail in the Methods section.

The most commonly used estimator with network surveys in current
practice is the VH estimator.  The other variations in use such as SH
are related to it and are based on the same assumptions plus the
additional assumption of a first-order Markov process in transitions
between node attribute values during the sampling.  The SH estimator
is used mainly for binary attribute variables.  For categorical
variables it has the property that the proportion estimates do not add
to one without extra adjusting of one kind or another, and it is not
well suited to continuous variables.  

\subsection{Mean degree and concurrency}

Among the most important quantities to estimate in relation to spread
of HIV are the means and proportions of link-related variables.  Two
of widespread interest are mean degree and concurrency.  Mean degree
is the average number of partners per person in the population.  The
most common definition of concurrency is the proportion of people in the
population who have two or more partners.  This and related
definitions of concurrency and their role in spreading HIV are
discussed in \cite{kretzschmar1996measures}, \cite{morris1997concurrent}, and
\cite{admiraal2016modeling}.  A high number for either of these is an
indication that an epidemic could spread rapidly in the population
once it starts there.  With the reference data set we are using for
the simulations, a link indicates either a sexual relationship, a drug
relationship, or a friendship relationship.  Because friendship is
included, the proportion of people with at least one relationship is
very close to 1.00.  Therefore we use a definition of concurrency
here, which might be called ``$k$-concurrency with k=10,'' where a person is
concurrent if they have more than 10 relationships.  The true
proportion of this is .82.  The purpose here is to compare estimation
methods for the mean of a characteristic of a node that is highly
related to links.  Because inclusion probabilities in network
sampling depend on the pattern of links in the population in
interaction with the design, these variables are the most sensitive to
choice of estimator.

The bars in Figure \ref{fig:bardegree} show the mean squared error of
three estimators for estimating mean degree with the regular snowball
design.  We include the sample mean as an estimator because it is the
one most commonly used to summarize the results in a snowball survey.
We include the estimator $\hat\mu_d = \sum [(y_i/d_i)/(1/d_i)]$, which
uses the self-reported degree $d_i$ as in place of the inclusion
probability $\pi_i$ in Brewer's estimator, because it is widely used,
though not recommended for snowball sampling.  It would be the
correct estimator, equal to Brewers, if the design used had been a
random walk instead of a snowball design, and if he self-reported
degrees were accurate.   It was used by \cite{salganik2004sampling} to
estimate mean degree and by 
\cite{volz2008probability} for estimating any kind of variable.  We
abbreviate it here VH.

Looking at the actual numbers represented in Figure
\ref{fig:bardegree}, with the snowball design the MSEs of y-bar, VH,
and the new estimator are respectively 40.453589,  7.249053, and 0.264755.
Dividing the MSE of the sample mean by that of the new estimator gives
thee relative efficiency of the new estimator as 153.  The relative
efficiency of the new estimator for the VH estimator is 29.  The high
MSE of the sample mean and VH for snowball sampling is almost entirely
due to bias.  For any estimator MSE = Variance + (Bias)$^2$.  For the
sample mean, the bias squared is over 99\% of the variance.  For the
VH the squared bias is  99\% of the variance.  So a bar plot of the
bias of the estimators looks almost exactly the same as the bar plot
for MSE in  Figure \ref{fig:bardegree}.

The large biases result from using the wrong sampling weights in the
estimators in relation to the actual inclusion probabilities $\pi_i$.
The new estimator estimates those inclusion probabilities using the
sample network of the data in interaction with the actual snowball
design used or a close approximation to it.  Because the estimated
inclusion probabilities $f_i$ come out close to proportional to the
actual inclusion probabilities $\pi_i$, the bias is virtually
eliminated with the new estimation method.

For the snowball design allowing re-recruiitments the relative
efficiency of the new estimator to y-bar for estimating mean degree is
329, and to VH the relative efficiency of the new estimator is 59.
The high MSEs of y-bar and VH again are mostly due to bias.

For the estimate of $k$-concurrency, with $k=10$, the MSE values with
the regular snowball design for y-bar, VH, and the new estimator are
0.010717  0.043369    0.000447.  The relative efficiency of the new
estimator to the sample mean is 24.  The relative efficiency of the
new estimator to the VH estimator is 97.  The squared bias accounts
for 
99\% of the MSE of y-bar, and 98\% of the MSE of the VH estimator.

For the snowball design allowing re-recruitments the relative
efficiency of the new estimator to y-bar for estimating
$k$-concurrency, with $k=10$,  is
17, and to VH the relative efficiency of the new estimator is 70.
The high MSEs of y-bar and VH again are explained largely by the bias
due to the incorrect weightings relative to the inclusion
probabilities of the snowball design used in the survey.  

Notice also that the magnitudes of MSE of y-bar and VH reverse in
relation to each other, with VH performing better than y-bar for mean
degree and worse than VH for $k$-concurrency.  When the sampling
weights of estimators differ significantly from the inclusion
probabilities of the survey, the estimators not only perform poorly
in most cases, but the performance is erratic, depending on the exact
configuration of values of variables of interest and of the sampling
weights in the population.  In the present study, only the new
estimator, with its inclusion probabilities estimated from the sample
data and the actual design, performed consistently well for every
variable and with each design.

\subsection{Attribute variables}

\begin{figure}
 \centerline{\includegraphics[width = 1.0\textwidth]{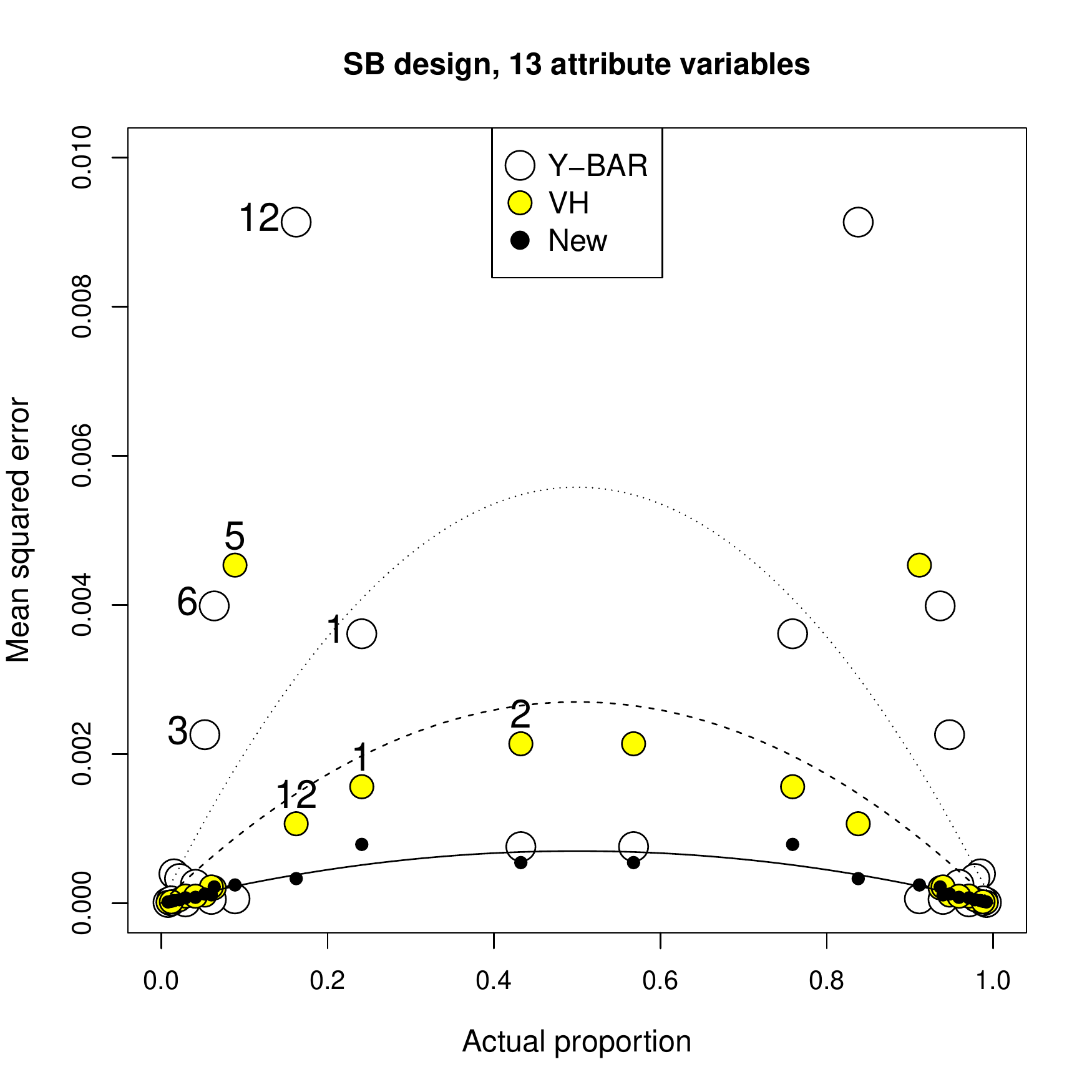}}
 \caption{With the regular snowball design, the mean squared error of
   three estimators of population proportion for each of 13 individual
   attributes. The three estimators are the sample mean (Y-BAR,
   white), weighting by reciprocal of self-reported degree (VH,
   yellow), and the new estimator (NEW, black).The new estimator
   (black) is compared to the current estimator (yellow).  To help see
   the pattern, the MSE for estimating the compliment of each
   attribute is also shown.  The compliment of sex work client, for
   example, is not-client.  A parabolic curve is fitted by weighted
   least squares to the MSEs of the new estimator (solid line) and the
   current estimator (dashed line).  The new estimator MSEs have the
   lowest fitted curve and the best fit.  }
\label{fig:parabolasb}
\end{figure}

\begin{figure}
 \centerline{\includegraphics[width =
   1.0\textwidth]{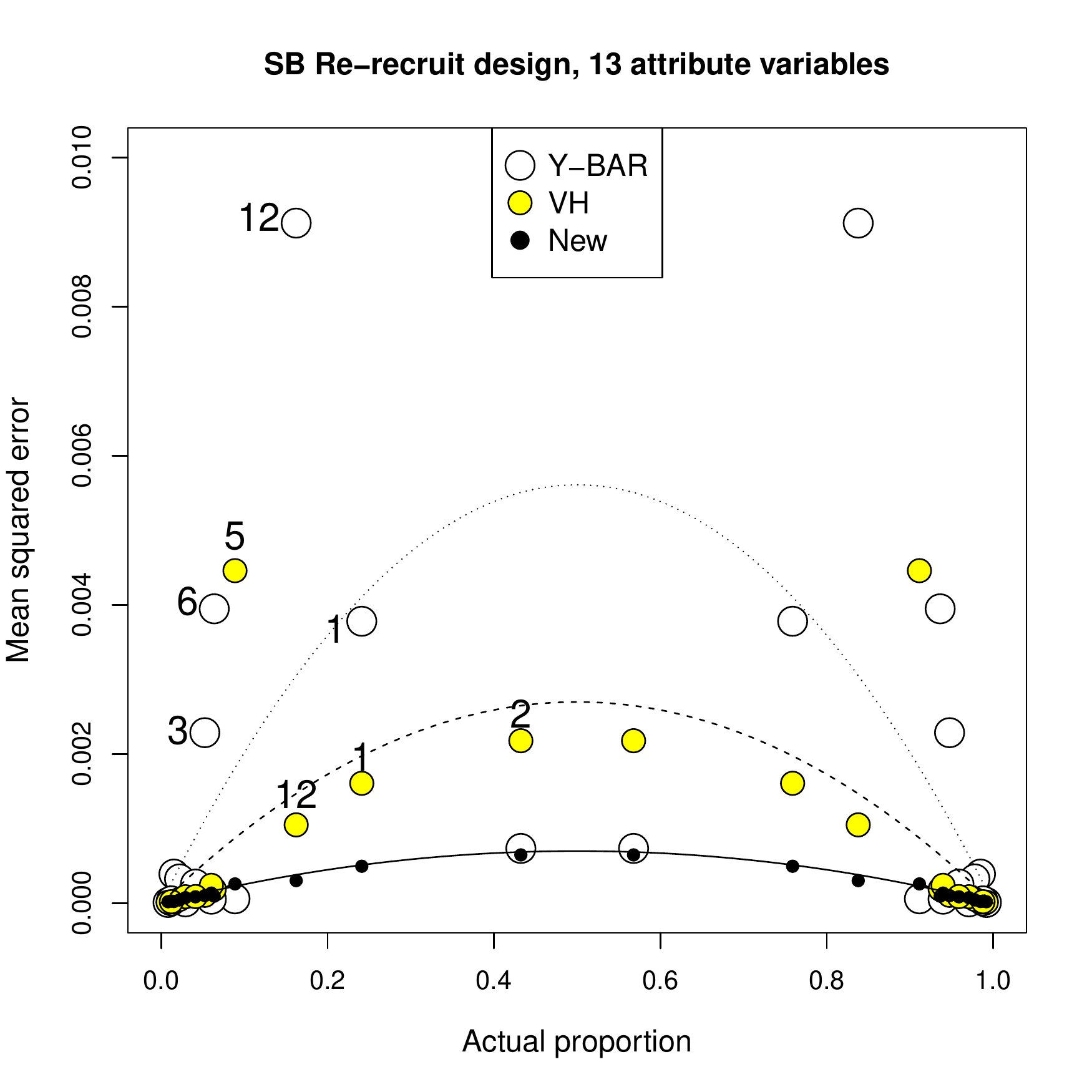}}
 \caption{With the snowball design allowing re-recruiitments, the mean
   squared error of three estimators of population proportion for each
   of 13 individual attributes. The three estimators are the sample
   mean (Y-BAR, white), weighting by reciprocal of self-reported
   degree (VH, yellow), and the new estimator (NEW, black).The new
   estimator (black) is compared to the current estimator (yellow).
   To help see the pattern, the MSE for estimating the compliment of
   each attribute is also shown.  The compliment of sex work client,
   for example, is not-client.  A parabolic curve is fitted by
   weighted least squares to the MSEs of the new estimator (solid
   line) and the current estimator (dashed line).  The new estimator
   MSEs have the lowest fitted curve and the best fit.  }
\label{fig:parabolasbfrr}
\end{figure}

The Colorado Springs Study node data includes 13 individual attribute variables
such as sex worker, client of sex worker, or unemployed.  These are
variables 1 through 13 in Table 1.  For an individual, the value is 1
if the individual has the attribute and 0 otherwise.  The object for
inference for each attribute is to estimate the proportion of people
in the population having that attribute.  Most of these variables,
such as sex or race or employment status, are not strongly or
consistently related to inclusion probabilities.  Still, for
design-based estimators to work well it helps to have the estimated
inclusion probabilities close to the actual inclusion probabilities.
For the simulations, missing values were arbitrarily set to zero so
that sample sizes would be the same for all variables.

%parabola

With a conventional simple random sampling design for estimating a
population proportion, the mean squared error of the estimate is a
parabola-shaped function of the actual population proportion.  The
actual proportion has to be between zero and one.  The MSE is highest
when the actual proportion is one-half and the MSE is zero when the
actual proportion is zero or one.  With the network designs and their
unequal inclusion probabilities the situation is more complex, but it
is still the case that the actual proportion has to be between zero
and one and that the MSE will be zero if the actual proportion is zero
or one.

To help see the pattern in the mean squared errors for estimating the
population proportions of the 13 attribute variables, the MSE for each
variable is plotted in Figure 3 against the actual proportion of
people having that attribute in the Colorado Springs study population.  For
each of the 13 variables, we can also estimate the proportion for its
complement.  The compliment of ``client'', for example, is ``not
client''.  The proportion for the compliment is 1 minus the proportion
with the attribute, and the MSE for estimation of the compliment is the
same as the the MSE for estimation of the original variable.  This
gives us 26 variables for each plot in Figure 3, with actual
proportions ranging from near 0 to near 1.  The original variables are
on the left, since the actual proportions are all less than one-half.
The compliment variables provide redundant information but clarify the
pattern in the MSEs.

The MSE with the new estimator (black in plots) is lower than that 
of the current estimator in all cases except for some of the ones with
actual proportion near to zero for which the MSE is very small with
either estimator.  While the MSEs of the new estimator fall
rather close to the fitted parabola (solid line), the MSEs of the
current estimator are more erratic and the fitted parabola (dashed
line) is higher.  The overall higher MSEs and erratic pattern with the
current estimator result from the discrepancies between actual
inclusion probabilities and those used in estimation.

% > print(xtable(data.frame(name=sbtable$name[1:15], actual=sbtable$actual[1:15], E.se = newcisbdata$E.se, halfwidth=newcisbdata$E.halfwidth, cerage=newcisbdata$coverage), caption='SB: Confidence Interval Coverage', label='SBcoverage', digits=c(2,2,2,3,2,2)), include.rownames = TRUE)
% latex table generated in R 3.6.3 by xtable 1.8-4 package
% Sun Jul 12 14:37:40 2020
\begin{table}[ht]
  \centering
  \caption{SB: Confidence Interval Coverage, median is .94}
\begin{tabular}{rlrrrr}
  \hline
 & name & actual & E.se & width & coverage \\ 
  \hline
1 & nonwhite & 0.24 & 0.023 & 0.04 & 0.91 \\ 
  2 & female & 0.43 & 0.029 & 0.06 & 0.98 \\ 
  3 & sex work & 0.05 & 0.010 & 0.02 & 0.94 \\ 
  4 & procures & 0.02 & 0.005 & 0.01 & 0.96 \\ 
  5 & sex-work client & 0.09 & 0.016 & 0.03 & 0.95 \\ 
  6 & sells drugs & 0.06 & 0.011 & 0.02 & 0.92 \\ 
  7 & makes drugs & 0.01 & 0.004 & 0.01 & 0.80 \\ 
  8 & stealing & 0.02 & 0.007 & 0.01 & 0.94 \\ 
  9 & retired & 0.03 & 0.009 & 0.02 & 0.92 \\ 
  10 & homemaker & 0.06 & 0.011 & 0.02 & 0.94 \\ 
  11 & disabled & 0.04 & 0.009 & 0.02 & 0.95 \\ 
  12 & unemployed & 0.16 & 0.017 & 0.03 & 0.96 \\ 
  13 & homeless & 0.01 & 0.005 & 0.01 & 0.86 \\ 
  14 & no. partners & 7.88 & 0.287 & 0.56 & 0.70 \\ 
  15 & concurrency & 0.82 & 0.035 & 0.07 & 1.00 \\ 
   \hline
\end{tabular}
\label{SBcoverage}
\end{table}
 
% > print(xtable(data.frame(name=sbtable$name[1:15], actual=sbtable$actual[1:15], E.se = newcisbfrrdata$E.se, halfwidth=newcisbfrrdata$E.halfwidth, cerage=newcisbfrrdata$coverage), caption='SB re-recruit design: Confidence Interval Coverage', label='SBcoverage', digits=c(2,2,2,3,2,2)), include.rownames = TRUE)
% latex table generated in R 3.6.3 by xtable 1.8-4 package
% Sun Jul 12 14:56:19 2020
\begin{table}[ht]
  \centering
  \caption{SB re-recruit design: Confidence Interval Coverage, median is .92} 
\begin{tabular}{rlrrrr}
  \hline
 & name & actual & E.se & width & coverage \\ 
  \hline
1 & nonwhite & 0.24 & 0.024 & 0.05 & 0.97 \\ 
  2 & female & 0.43 & 0.031 & 0.06 & 0.98 \\ 
  3 & sex work & 0.05 & 0.010 & 0.02 & 0.93 \\ 
  4 & procures & 0.02 & 0.005 & 0.01 & 0.81 \\ 
  5 & sex-work client & 0.09 & 0.016 & 0.03 & 0.95 \\ 
  6 & sells drugs & 0.06 & 0.011 & 0.02 & 0.95 \\ 
  7 & makes drugs & 0.01 & 0.004 & 0.01 & 0.75 \\ 
  8 & stealing & 0.02 & 0.007 & 0.01 & 0.90 \\ 
  9 & retired & 0.03 & 0.009 & 0.02 & 0.91 \\ 
  10 & homemaker & 0.06 & 0.012 & 0.02 & 0.92 \\ 
  11 & disabled & 0.04 & 0.010 & 0.02 & 0.92 \\ 
  12 & unemployed & 0.16 & 0.018 & 0.03 & 0.95 \\ 
  13 & homeless & 0.01 & 0.005 & 0.01 & 0.80 \\ 
  14 & no. partners & 7.88 & 0.267 & 0.52 & 0.85 \\ 
  15 & concurrency & 0.82 & 0.038 & 0.07 & 0.99 \\ 
   \hline
\end{tabular}
\label{SBfrrcoverage}
\end{table}

The parabolas in the plots have form ${\rm MSE}=ap(1-p)$ where $p$ is the
actual proportion, which is known for each of the 13 attribute
variables in the simulation population. The coefficient $a$ measures
the height of the fitted parabola for a given estimator-design
combination.  Since the relationship is linear with increasing variance in the
quantity $p(1-p)$, the weighted least squares regression estimator of
the coefficient $a$ is a simple ratio estimator.  Consider that the
variable $y$ = MSE linearly increases with the variable $x = p(1-p)$
and that the variance of $y$ about this line increases with $x$,
approximately linearly.  From weighted regression results this
suggests, as an estimator for the slope $a$, the ratio estimator $hat
a = \sum y_i / \sum x_i$.  So for an overall comparison of the
estimator y-bar with the new estimator, let $a_{\rm ybar}$ and
$a_{\rm new}$ represent the estimated parameters of the respective lines.
The ratio $a_{\rm ybar} /    a_{\rm new}$ is simply  the sum
of the MSEs with y-bar divided by the sum of the MSEs with the new
estimator, because the $x$-values are the same with each design, so
their sums divide out.  Thus the ratio of the estimated parabola
parameters $a$ is simply the ratio of the average MSE for the two
designs.  It's also the ratio of heights of the two parabolas at any
point.  This ratio provides a measure of average relative efficiency
of the the two designs.

For the regular snowball design, the overall relative efficiency of the
new estimator to the sample mean is $a_{\rm ybar} /    a_{\rm new} =
 0.02231997 /  0.002791733 = 8.0$.  The relative efficiency of the new
 estimator to the VH estimator is  0.01079285 /  0.002791733 = 3.9.

 For the snowball design allowing re-recruitments, the overall relative efficiency of the
 new estimator to the sample meean is $a_{\rm ybar} /    a_{\rm new}
 = 0.02244716 /  0.002445438 = 9.2$.
The relative efficiency of the new
estimator to the VH estimator is  0.01078216 /  0.002445438 = 4.4.  To
see these relative parabola heights in Figures 5 and 6, note the position of
zero on the vertical axis.

% new sb ci table

Confidence interval coverage probabilities for each variable for each
of the 15 variables are given in Tables 1 and 2.  The median coverage
probability of the intervals for the 15 characteristics estimated with
each of two designs is .94.

\section{Discussion and conclusions}

Snowball designs are the most natural of the network designs and have
many desirable properties.  They self-allocate so that most of the
sample size goes into the most highly connected parts of the
population.  This leads investigators to those areas where risk of
spread of a virus is highest.  This makes snowball designs highly
effective for distributing an intervention such as a vaccine to a
population.

Snowball designs have been used less than they should be because a
simple design-based estimator has not been available.  The new
method of this paper provide such estimators.

Snowball designs and accurate estimators to go with them are needed
for preventing new pandemics from the coronaviruses family of viruses.
They are needed for understanding sexual networks on which HIV spreads
and opioid misuse networks which spread their own harms and
occasionally provide explosive terrain for HIV.  The new estimators
can also be used for adaptive spatial designs for hard-to-sample
animal species, by translating the spatial structure into network
form.

Accurate estimation with snowball designs requires estimating the
inclusion probabilities $\pi_i$ from the data taking the actual survey
sampling design into account.  Sample means and proportions do not
provide good estimates or data summaries for snowball samples.
Estimators that weight observations by the reciprocal of the unit's
degree also do not work well with snowball sampling designs. 

The new estimates are fast to compute, scale up well, and are based on the
data and the design actually used rather than unrealistic
assumptions.  They are low-bias, accurate, and give reliable
confidence intervals.

%  The \backmatter command formats the subsequent headings so that they
%  are in the journal style.  Please keep this command in your document
%  in this position, right after the final section of the main part of 
%  the paper and right before the Acknowledgements, Supporting Information (Supplementary %  Materials),   and References sections. 

%\backmatter

%  This section is optional.  Here is where you will want to cite
%  grants, people who helped with the paper, etc.  But keep it short!

\section*{Acknowledgements}

  This research was supported by Natural Science and Engineering
  Research Council of Canada (NSERC) Discovery grant RGPIN327306.  I
  would like to thank John Potterat and Steve Muth for making the
  Colorado Springs study data available and for their generous help
  explaining it. I would like to express appreciation for the
  participants in that study who shared their personal information
  with the researchers so that it could be made available in
  anonymized form to the research community and contribute to a
  solution to HIV and and addiction epidemics and to basic
  understanding of social networks.
\vspace*{-8pt}

%  Here, we create the bibliographic entries manually, following the
%  journal style.  If you use this method or use natbib, PLEASE PAY
%  CAREFUL ATTENTION TO THE BIBLIOGRAPHIC STYLE IN A RECENT ISSUE OF
%  THE JOURNAL AND FOLLOW IT!  Failure to follow stylistic conventions
%  just lengthens the time spend copyediting your paper and hence its
%  position in the publication queue should it be accepted.

%  We greatly prefer that you incorporate the references for your
%  article into the body of the article as we have done here 
%  (you can use natbib or not as you choose) than use BiBTeX,
%  so that your article is self-contained in one file.
%  If you do use BiBTeX, please use the .bst file that comes with 
%  the distribution.  In this case, replace the thebibliography
%  environment below by 
%
% use this when using biblliography{sbrefs} for the bib file
  \bibliographystyle{apalike} 
% \bibliography{mybibilo.bib}

% \begin{thebibliography}{}

\bibliography{sbrefs}

\begin{thebibliography}{}

\bibitem[Admiraal et~al., 2016]{admiraal2016modeling}
Admiraal, R., Handcock, M.~S., et~al. (2016).
\newblock Modeling concurrency and selective mixing in heterosexual partnership
  networks with applications to sexually transmitted diseases.
\newblock {\em The Annals of Applied Statistics}, 10(4):2021--2046.

\bibitem[Andersen et~al., 2020]{andersen2020proximal}
Andersen, K.~G., Rambaut, A., Lipkin, W.~I., Holmes, E.~C., and Garry, R.~F.
  (2020).
\newblock The proximal origin of sars-cov-2.
\newblock {\em Nature medicine}, 26(4):450--452.

\bibitem[Atkinson and Flint, 2001]{atkinson2001accessing}
Atkinson, R. and Flint, J. (2001).
\newblock Accessing hidden and hard-to-reach populations: Snowball research
  strategies.
\newblock {\em Social research update}, 33(1):1--4.

\bibitem[Baraff et~al., 2016]{baraff2016estimating}
Baraff, A.~J., McCormick, T.~H., and Raftery, A.~E. (2016).
\newblock Estimating uncertainty in respondent-driven sampling using a tree
  bootstrap method.
\newblock {\em Proceedings of the National Academy of Sciences},
  113(51):14668--14673.

\bibitem[Birnbaum and Sirken, 1965]{birnbaum1965design}
Birnbaum, Z. and Sirken, M.~G. (1965).
\newblock Design of sample surveys to estimate the prevalence of rare diseases:
  three unbiased estimates, vital and health statistics, series 2.
\newblock {\em Government Printing Office, Washington, DC}.

\bibitem[Brewer, 1963]{brewer1963ratio}
Brewer, K. (1963).
\newblock Ratio estimation and finite populations: Some results deducible from
  the assumption of an underlying stochastic process.
\newblock {\em Australian Journal of Statistics}, 5(3):93--105.

\bibitem[Campbell et~al., 2017]{campbell2017detailed}
Campbell, E.~M., Jia, H., Shankar, A., Hanson, D., Luo, W., Masciotra, S.,
  Owen, S.~M., Oster, A.~M., Galang, R.~R., Spiller, M.~W., et~al. (2017).
\newblock Detailed transmission network analysis of a large opiate-driven
  outbreak of hiv infection in the united states.
\newblock {\em The Journal of infectious diseases}, 216(9):1053--1062.

\bibitem[Chow and Thompson, 2003a]{chow2003bayesian}
Chow, M. and Thompson, S.~K. (2003a).
\newblock A bayesian approach to estimation with link-tracing sampling designs.
\newblock {\em Survey Methodology}, 29:197--205.

\bibitem[Chow and Thompson, 2003b]{chow2003estimation}
Chow, M. and Thompson, S.~K. (2003b).
\newblock Estimation with link-tracing sampling designs a bayesian approach.
\newblock {\em Survey Methodology}, 29(2):197--206.

\bibitem[Corman et~al., 2015]{corman2015evidence}
Corman, V.~M., Baldwin, H.~J., Tateno, A.~F., Zerbinati, R.~M., Annan, A.,
  Owusu, M., Nkrumah, E.~E., Maganga, G.~D., Oppong, S., Adu-Sarkodie, Y.,
  et~al. (2015).
\newblock Evidence for an ancestral association of human coronavirus 229e with
  bats.
\newblock {\em Journal of virology}, 89(23):11858--11870.

\bibitem[Frank, 1977]{frank1977survey}
Frank, O. (1977).
\newblock Survey sampling in graphs.
\newblock {\em Journal of Statistical Planning and Inference}, 1(3):235--264.

\bibitem[Frank and Snijders, 1994]{frank1994estimating}
Frank, O. and Snijders, T. (1994).
\newblock Estimating the size of hidden populations using snowball sampling.
\newblock {\em Journal of Official Statistics}, 10:53--53.

\bibitem[Ge et~al., 2013]{ge2013isolation}
Ge, X.-Y., Li, J.-L., Yang, X.-L., Chmura, A.~A., Zhu, G., Epstein, J.~H.,
  Mazet, J.~K., Hu, B., Zhang, W., Peng, C., et~al. (2013).
\newblock Isolation and characterization of a bat sars-like coronavirus that
  uses the ace2 receptor.
\newblock {\em Nature}, 503(7477):535--538.

\bibitem[Gile, 2011]{gile2011improved}
Gile, K.~J. (2011).
\newblock Improved inference for respondent-driven sampling data with
  application to hiv prevalence estimation.
\newblock {\em Journal of the American Statistical Association},
  106(493):135--146.

\bibitem[Graham and Baric, 2010]{graham2010recombination}
Graham, R.~L. and Baric, R.~S. (2010).
\newblock Recombination, reservoirs, and the modular spike: mechanisms of
  coronavirus cross-species transmission.
\newblock {\em Journal of virology}, 84(7):3134--3146.

\bibitem[Handcock and Gile, 2010]{handcock2010modeling}
Handcock, M. and Gile, K. (2010).
\newblock Modeling social networks from sampled data.
\newblock {\em The Annals of Applied Statistics}, 4(1):5--25.

\bibitem[Handcock and Gile, 2007]{handcock2007modeling}
Handcock, M.~S. and Gile, K. (2007).
\newblock Modeling social networks with sampled or missing data.
\newblock {\em \textrm Center for statistics and the social sciences working
  paper no.75. University of Washington, Department of Statistics.}

\bibitem[Handcock and Gile, 2011]{handcock2011comment}
Handcock, M.~S. and Gile, K.~J. (2011).
\newblock Comment: On the concept of snowball sampling.
\newblock {\em Sociological Methodology}, 41(1):367--371.

\bibitem[Heckathorn, 1997]{heckathorn1997respondent}
Heckathorn, D.~D. (1997).
\newblock Respondent-driven sampling: a new approach to the study of hidden
  populations.
\newblock {\em Social problems}, 44(2):174--199.

\bibitem[Heckathorn, 2007]{heckathorn20076}
Heckathorn, D.~D. (2007).
\newblock Extensions of respondent-driven sampling: Analyzing continuous
  variables and controlling for differential recruitment.
\newblock {\em Sociological Methodology}, 37(1):151--208.

\bibitem[Heckathorn, 2011]{doi:10.1111/j.1467-9531.2011.01244.x}
Heckathorn, D.~D. (2011).
\newblock Comment: Snowball versus respondent-driven sampling.
\newblock {\em Sociological Methodology}, 41(1):355--366.
\newblock PMID: 22228916.

\bibitem[Hu et~al., 2017]{hu2017discovery}
Hu, B., Zeng, L.-P., Yang, X.-L., Ge, X.-Y., Zhang, W., Li, B., Xie, J.-Z.,
  Shen, X.-R., Zhang, Y.-Z., Wang, N., et~al. (2017).
\newblock Discovery of a rich gene pool of bat sars-related coronaviruses
  provides new insights into the origin of sars coronavirus.
\newblock {\em PLoS pathogens}, 13(11):e1006698.

\bibitem[Huong et~al., 2020]{huong2020journal.pone.0237129}
Huong, N.~Q., Nga, N. T.~T., Long, N.~V., Luu, B.~D., Latinne, A., Pruvot, M.,
  Phuong, N.~T., Quang, L. T.~V., Hung, V.~V., Lan, N.~T., Hoa, N.~T., Minh,
  P.~Q., Diep, N.~T., Tung, N., Ky, V.~D., Roberton, S.~I., Thuy, H.~B., Long,
  N.~V., Gilbert, M., Wicker, L., Mazet, J. A.~K., Johnson, C.~K., Goldstein,
  T., Tremeau-Bravard, A., Ontiveros, V., Joly, D.~O., Walzer, C., Fine, A.~E.,
  and Olson, S.~H. (2020).
\newblock Coronavirus testing indicates transmission risk increases along
  wildlife supply chains for human consumption in viet nam, 2013-2014.
\newblock {\em PLOS ONE}, 15(8):1--20.

\bibitem[Kretzschmar and Morris, 1996]{kretzschmar1996measures}
Kretzschmar, M. and Morris, M. (1996).
\newblock Measures of concurrency in networks and the spread of infectious
  disease.
\newblock {\em Mathematical biosciences}, 133(2):165--195.

\bibitem[Latinne et~al., 2020]{latinne2020origin}
Latinne, A., Hu, B., Olival, K.~J., Zhu, G., Zhang, L., Li, H., Chmura, A.~A.,
  Field, H.~E., Zambrana-Torrelio, C., Epstein, J.~H., et~al. (2020).
\newblock Origin and cross-species transmission of bat coronaviruses in china.
\newblock {\em bioRxiv}.

\bibitem[Li et~al., 2019]{li2019human}
Li, H., Mendelsohn, E., Zong, C., Zhang, W., Hagan, E., Wang, N., Li, S., Yan,
  H., Huang, H., Zhu, G., et~al. (2019).
\newblock Human-animal interactions and bat coronavirus spillover potential
  among rural residents in southern china.
\newblock {\em Biosafety and Health}, 1(2):84--90.

\bibitem[Li et~al., 2005]{li2005bats}
Li, W., Shi, Z., Yu, M., Ren, W., Smith, C., Epstein, J.~H., Wang, H., Crameri,
  G., Hu, Z., Zhang, H., et~al. (2005).
\newblock Bats are natural reservoirs of sars-like coronaviruses.
\newblock {\em Science}, 310(5748):676--679.

\bibitem[Montgomery and Macdonald, 2020]{montgomery2020covid}
Montgomery, R.~A. and Macdonald, D.~W. (2020).
\newblock Covid-19, health, conservation, and shared wellbeing: Details matter.
\newblock {\em Trends in Ecology \& Evolution}.

\bibitem[Morris and Kretzschmar, 1997]{morris1997concurrent}
Morris, M. and Kretzschmar, M. (1997).
\newblock Concurrent partnerships and the spread of hiv.
\newblock {\em Aids}, 11(5):641--648.

\bibitem[Olival et~al., 2017a]{olival2017erratum}
Olival, K.~J., Hosseini, P.~R., Zambrana-Torrelio, C., Ross, N., Bogich, T.~L.,
  and Daszak, P. (2017a).
\newblock Erratum: host and viral traits predict zoonotic spillover from
  mammals.
\newblock {\em Nature}, 548(7669):612--612.

\bibitem[Olival et~al., 2017b]{olival2017host}
Olival, K.~J., Hosseini, P.~R., Zambrana-Torrelio, C., Ross, N., Bogich, T.~L.,
  and Daszak, P. (2017b).
\newblock Host and viral traits predict zoonotic spillover from mammals.
\newblock {\em Nature}, 546(7660):646--650.

\bibitem[Pattison et~al., 2013]{pattison2013conditional}
Pattison, P.~E., Robins, G.~L., Snijders, T.~A., and Wang, P. (2013).
\newblock Conditional estimation of exponential random graph models from
  snowball sampling designs.
\newblock {\em Journal of mathematical psychology}, 57(6):284--296.

\bibitem[Peters et~al., 2016]{peters2016hiv}
Peters, P.~J., Pontones, P., Hoover, K.~W., Patel, M.~R., Galang, R.~R.,
  Shields, J., Blosser, S.~J., Spiller, M.~W., Combs, B., Switzer, W.~M.,
  et~al. (2016).
\newblock Hiv infection linked to injection use of oxymorphone in indiana,
  2014--2015.
\newblock {\em New England Journal of Medicine}, 375(3):229--239.

\bibitem[Potterat et~al., 1999]{potterat1999network}
Potterat, J.~J., Rothenberg, R.~B., and Muth, S.~Q. (1999).
\newblock Network structural dynamics and infectious disease propagation.
\newblock {\em International journal of STD \& AIDS}, 10(3):182--185.

\bibitem[Rohe et~al., 2019]{rohe2019critical}
Rohe, K. et~al. (2019).
\newblock A critical threshold for design effects in network sampling.
\newblock {\em The Annals of Statistics}, 47(1):556--582.

\bibitem[Salganik and Heckathorn, 2004]{salganik2004sampling}
Salganik, M.~J. and Heckathorn, D.~D. (2004).
\newblock Sampling and estimation in hidden populations using respondent-driven
  sampling.
\newblock {\em Sociological methodology}, 34(1):193--240.

\bibitem[Snijders, 1992]{snijders1992estimation}
Snijders, T.~A. (1992).
\newblock Estimation on the basis of snowball samples: how to weight?
\newblock {\em Bulletin of Sociological Methodology/Bulletin de
  M{\'e}thodologie Sociologique}, 36(1):59--70.

\bibitem[Spreen, 1992]{spreen1992rare}
Spreen, M. (1992).
\newblock Rare populations, hidden populations, and link-tracing designs: What
  and why?
\newblock {\em Bulletin of Sociological Methodology/Bulletin de Methodologie
  Sociologique}, 36(1):34--58.

\bibitem[Tang et~al., 2020]{tang2020origin}
Tang, X., Wu, C., Li, X., Song, Y., Yao, X., Wu, X., Duan, Y., Zhang, H., Wang,
  Y., Qian, Z., et~al. (2020).
\newblock On the origin and continuing evolution of sars-cov-2.
\newblock {\em National Science Review}.

\bibitem[Thompson, 2011]{thompson2011adaptive}
Thompson, S. (2011).
\newblock Adaptive network and spatial sampling.
\newblock {\em Survey Methodology}, 37(2):183--196.

\bibitem[Thompson, 1990]{thompson1990adaptive}
Thompson, S.~K. (1990).
\newblock Adaptive cluster sampling.
\newblock {\em Journal of the American Statistical Association},
  85(412):1050--1059.

\bibitem[Thompson, 2006]{thompson2006aws}
Thompson, S.~K. (2006).
\newblock Adaptive web sampling.
\newblock {\em Biometrics}, 62(4):1224--1234.

\bibitem[Thompson, 2017]{thompson2017adaptive}
Thompson, S.~K. (2017).
\newblock Adaptive and network sampling for inference and interventions in
  changing populations.
\newblock {\em Journal of Survey Statistics and Methodology}, 5(1):1--21.

\bibitem[Thompson and Collins, 2002]{thompson2002adaptive}
Thompson, S.~K. and Collins, L.~M. (2002).
\newblock Adaptive sampling in research on risk-related behaviors.
\newblock {\em Drug and Alcohol Dependence}, 68:57--67.

\bibitem[Thompson and Frank, 2000]{thompson2000model}
Thompson, S.~K. and Frank, O. (2000).
\newblock Model-based estimation with link-tracing sampling designs.
\newblock {\em Survey methodology}, 26(1):87--98.

\bibitem[Volz and Heckathorn, 2008]{volz2008probability}
Volz, E. and Heckathorn, D.~D. (2008).
\newblock Probability based estimation theory for respondent driven sampling.
\newblock {\em Journal of official statistics}, 24(1):79.

\bibitem[Wang and Anderson, 2019]{wang2019viruses}
Wang, L.-F. and Anderson, D.~E. (2019).
\newblock Viruses in bats and potential spillover to animals and humans.
\newblock {\em Current opinion in virology}, 34:79--89.

\bibitem[Wang et~al., 2006]{wang2006review}
Wang, L.-F., Shi, Z., Zhang, S., Field, H., Daszak, P., and Eaton, B.~T.
  (2006).
\newblock Review of bats and sars.
\newblock {\em Emerging infectious diseases}, 12(12):1834.

\bibitem[Zhou et~al., 2020]{zhou2020pneumonia}
Zhou, P., Yang, X.-L., Wang, X.-G., Hu, B., Zhang, L., Zhang, W., Si, H.-R.,
  Zhu, Y., Li, B., Huang, C.-L., et~al. (2020).
\newblock A pneumonia outbreak associated with a new coronavirus of probable
  bat origin.
\newblock {\em Nature}, 579(7798):270--273.

\end{thebibliography}

% \begin{thebibliography}{99}

% \bibitem[Kopka and Daly(2003)]{R1}
% Kopka~H and Daly~PW (2003) \textit{A Guide to \LaTeX}, 4th~edn.
% Addison-Wesley.

% % \bibitem[Lamport(1994)]{R2}
% Lamport~L (1994) \textit{\LaTeX: a Document Preparation System},
% 2nd~edn. Addison-Wesley.

% \end{thebibliography}

%  If your paper refers to supporting web material, then you MUST
%  include this section!!  See Instructions for Authors at the journal
%  website http://www.biometrics.tibs.org

%\section*{Supporting Information}

%Web supporting details and tables, referenced in Section 1, are
%available with this paper at the Biometrics website on Wiley Online Library.

\end{document}